\documentclass[aps,prd,twocolumn,floatfix]{revtex4}
\usepackage{graphicx}
\usepackage{dcolumn}
\usepackage{bm}
\usepackage{amsmath}
\begin{document}

\def\n{\noindent}
\def\be{\begin{equation}}    \def\ee{\end{equation}}
\def\bd{\begin{displaymath}} \def\ed{\end{displaymath}}
\def\ba{\begin{eqnarray}}    \def\ea{\end{eqnarray}}

\def\d{\rm d}
\def\A{\rm A}
\def\G{\rm G}
\def\H{\rm H}
\def\I{\rm I}
\def\N{\rm N}
\def\P{\rm P}
\def\R{\rm R}
\def\Rec{\rm Rec}
\def\S{\rm S}
\def\StoI{\S\to \I}
\def\ItoR{\I\to \R}
\def\eff{e\!f\!f}
\def\k{$k$}
\def\M{$M$}

\title{\large
The SHIR Model: Realistic Fits to COVID-19 Case Numbers}

\author{
T.Barnes\footnote{Email: tbarnes@utk.edu}
}

\affiliation{
University of Tennessee\\
Department of Physics and Astronomy\\
Knoxville, TN 37996-1200, USA
}

\date{\today}

\begin{abstract}
We consider a global (location independent) model of pandemic growth 
which generalizes the SIR model to accommodate important features of the COVID-19 
pandemic, notably the implementation of pandemic reduction measures. 
This ``SHIR" model is applied to COVID-19 data, and shows promise as a simple, 
tractable formalism with few parameters that can be used to model pandemic case numbers. 
As an example we show that the average time dependence of new COVID-19 cases per day 
from 15 Central and Western European countries is in good agreement with the analytic, 
parameter-free prediction of the model. 

\end{abstract}

\maketitle

\section{Epidemic Models}
\subsection{SIR Models}

In the models typically employed in epidemiology 
to follow the evolution of a pandemic \cite{Roc2014},
one begins by partitioning a total population of N individuals 
into a complete set of disjoint subsets, according to their medical histories. 
The simplest models of this type have only a few categories, such as 
1) the susceptible but as yet uninfected population S, 2) the infected but not recovered 
population I, who can infect S, and 3) the recovered population R, 
who can no longer infect S, and can themselves no longer be infected. 
First-order ordinary differential equations (ODEs) in time are assumed to
describe the coupling between these groups.
Once the appropriate initial conditions are specified, these ODEs can be integrated, 
which gives predictions for the subsequent evolution of the group populations.   

In this example, given constant rates $\{ r_{ij} \}$ for the two assumed transitions, 
$\StoI$ and $\ItoR$, the resulting ODEs for this simple ``SIR" model are 

\ba
&d{\S}/dt    &=   - r_{\StoI} \S                 \cr
\cr
&d{\I}/dt    &=   + r_{\StoI} \S - r_{\ItoR} \I   \cr
\cr
&d{\R}/dt    &=                 + r_{\ItoR} \I  
\label{eq:SIR_model}
\ea
We will refer to these two rate coefficients as 
$r_{\StoI} = r_{\I}$ and $r_{\ItoR} = r_{\R}$.
On adding these equations we find that 
\be
d{\S}/dt + d{\I}/dt + d{\R}/dt = d{\N}/dt = 0 \ ,
\label{eq:N_const}
\ee
so the total number of individuals is a constant, $\N_0$.

Since these equations are linear and homogeneous, it is convenient to divide by $\N_0$ 
and solve for the fraction of individuals in each category, such as $f_{\S} = {\S}/\N_0$. 
R or $f_{\R}$ can be inferred from the other two populations, using 
$\R = \N_0 - \S - \I$, or equivalently $f_{\R} = 1 - f_{\S} - f_{\I}$, so we need only 
solve the set (\ref{eq:SI_DEs}),

\ba
&df_{\S}/dt    &=   - r_{\I} f_{\S}     \ ,            \cr
\cr
&df_{\I}/dt    &=   + r_{\I} f_{\S} - r_{\R} f_{\I}   \ .
\label{eq:SI_DEs}
\ea
We will assume that the entire population is initially in category S 
(susceptible but not infected or recovered), so that $f_{\S}(t=0) = 1$ and
$f_{\I}(t=0) = 0$. Given these initial conditions we can solve the 
equations (\ref{eq:SI_DEs}), which also imply $f_{\R}$ through 
$f_{\R} = 1 - f_{\S} - f_{\I}$, with the results 
\ba
&f_{\S}(t) &= e^{-r_{\I} t}                                            \cr
\cr
&f_{\I}(t) &= \frac{r_{\I}}{r_{\R} -r_{\I} } \Big(e^{-r_{\I} t} - e^{-r_{\R} t}\Big) \cr 
\cr
&f_{\R}(t) &= 1 - \frac{1}{r_{\R} -r_{\I} } \Big(r_{\R} e^{-r_{\I} t} - r_{\I} e^{-r_{\R} t} \Big) . 
\label{eq:SIR_soln}
\ea

\begin{figure}[t]
\includegraphics[width=0.9\linewidth]{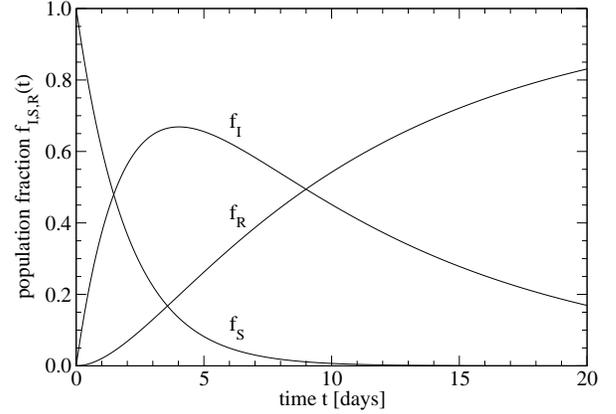}
\caption{An example of a solution of the SIR equations 
(\ref{eq:SI_DEs}), 
starting with the initial conditions that the fraction of susceptibles is 
$f_{\S} = 1.0$ (all), 
and the infection rate 
$r_{\I} = 0.5$~[days$^{-1}$] 
is five times the recovery rate, 
$r_{\R} = 0.1$~[days$^{-1}$].
}
\label{fig:SIR_eg}
\end{figure}
\eject

The behavior of this system is just what one would expect from the dynamics described in 
the SIR-type defining equations (\ref{eq:SIR_model}): The initial population of susceptibles 
($\S$)
decays exponentially at the specified rate $r_{\I}$ as they transition into infected
($\I$); 
the infected fraction increases to a maximum \cite{fI_max}, and then declines as it populates 
the final, recovered population ($\R$).
As $t\to\infty$ one asymptotically approaches a state in which all of the population 
has transitioned to recovered, $f_{\R}(t=+\infty) = 1$. 
As anticipated, at any intermediate time $0 < t < +\infty $, 
$f_{\S}(t) + f_{\I}(t) + f_{\I}(t) = 1$. 

As a specific example, the time dependence predicted by this SIR model is shown in Fig.\ref{fig:SIR_eg}, 
for $r_{\I} = 0.5$~[days$^{-1}$] and $r_{\R} = 0.1$~[days$^{-1}$] (an infection rate of five times 
the recovery rate). 

\subsection{A ``SHIR" Model for COVID-19}

Two important modifications to this simple pedagogical SIR model appear appropriate for 
a more realistic description of the COVID-19 pandemic. One is that the rate of transitions 
from susceptible (S) to infected (I) should be proportional to the product of these 
two populations, since their interaction determines this rate. 
This modifies the second evolution equation for the population fraction $f_{\I}(t)$ to
\be
df_{\I}/dt    =   + r_{\I} f_{\S}f_{\I}  - r_{\R} f_{\I}   \ .
\label{eq:SHIR_DE2}
\ee
This ``second-order reaction kinetics" in the infection process is 
a standard assumption in epidemic models \cite{Roc2014,Ker1927}.

A second, more novel modification of the SIR model is to
incorporate the effects of a social response to a high-profile 
pandemic such as COVID-19. In this case there has been strong encouragement from governments 
and medical authorities to reduce the growth of the pandemic by implementing social distancing 
and related measures. In this SHIR model these actions reduce the susceptible population 
by transferring them to a new category of ``hidden" individuals (H) who are not susceptible to
infection, which is populated from the original group of susceptibles (S) through another rate 
process,
\be
df_{\H}/dt    =   + r_{\H} f_{\S} \ . 
\label{eq:SHIR_DE0}
\ee

The individuals in ${\H}$ are not only the physically sequestered; simply changing behavior, 
such as avoiding public assemblies, crowded schools and public transportation, implementing rigorous 
hygienic practices, and many other measures contribute to the continued reduction of the effective 
susceptible population fraction $f_{\S}$ \cite{Hbox}. 

We consider $\tau = 1/r_{\H}$, the ``social response time," to be a measure of 
how quickly a society responds to a developing pandemic threat. We shall subsequently find that there
are interesting differences between countries in the fitted values of this parameter.

The partner DE for $f_{\S}$, similarly modified to include a second-order 
infection rate process as well as the new first-order social response transition term for ${\S}\to{\H}$, is
\be
df_{\S}/dt    =   - r_{\I} f_{\S}f_{\I} - r_{\H} f_{\S} \ .
\label{eq:SHIR_DE1}
\ee
The final ``recovery" DE for $d{\R}/dt$ is unchanged from the SIR model formula in (\ref{eq:SIR_model}). 
However since the ${\R}$ cases in practice include mortality as well as recovery, we will instead refer 
to the ${\R}$ population as ``resolved"~\cite{Ard}. 

\begin{figure}[t]
\vskip -5mm
\includegraphics[width=0.9\linewidth]{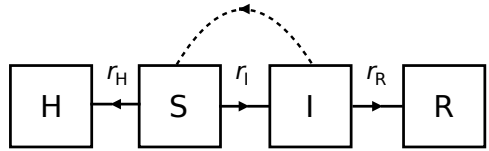}
\vskip -1cm
\caption{A block diagram of the ``SHIR" model, which generalizes SIR by introducing 
a ``hidden" population H and an ${{\S} \to {\H}}$ rate constant $r_{\H}$.}
\label{fig:SHIR_blockdiag1}
\end{figure}

Collecting these results gives our complete set of SHIR model DEs,
\ba
&df_{\H}/dt    &=   + r_{\H} f_{\S}                       \cr
\cr
&df_{\S}/dt    &=   - r_{\I} f_{\S}f_{\I} - r_{\H} f_{\S}    \cr
\cr
&df_{\I}/dt    &=   + r_{\I} f_{\S}f_{\I}  - r_{\R} f_{\I}   \cr
\cr
&df_{\R}/dt    &=                       + r_{\R} f_{\I}   \ .  
\label{eq:SHIR_DEs}
\ea 
Note that one recovers the three original pandemic DEs of Kermack~{\it et al.}~\cite{Ker1927} 
(their Eqs.(29)) through the substitutions 
$f_{(\S,\I,\R)} \to (x,y,z)/{\N}$ and $(r_{\I},r_{\R})\to (\kappa,\ell)$, and
removing the ``hidden population" of the SHIR model by setting $r_{\H}=0$
and $f_{\H}=0$. 

We note in passing that the infected (and still infective) individuals (all those in box ${\I}$) 
are also referred to in the literature as the  ``active" cases, numbering ${\A}(t)$. 
Here, ${\I}(t)$ and ${\A}(t)$ are identical; any individuals in box ${\I}$ who are no longer 
infective are immediately reclassified as ``resolved," and moved to box {\R}.

Since (\ref{eq:SHIR_DEs}) is a {\it nonlinear} set of ODEs 
for the four population fractions $f_{\S,\H,\I,\R}$, 
an exact analytical solution will not be possible, except in certain 
limits and approximations. 
However some important results can still be proven for this nonlinear system. 
One such result, obtained by adding the 4 equations in (\ref{eq:SHIR_DEs}), is that
the sum of the four populations fractions $f$ is a constant of motion. Thus
$f_{\H} + f_{\S} + f_{\I} + f_{\R} = 1$, or ${\H + \S + \I + \R}$ remains a constant, 
${\N}_0$.

Another useful result is that the population fractions of the two groups
H ``hidden" and R ``resolved" can be expressed simply in terms of 
the susceptible (S) and infected (I) fractions $f_{\S}$ and $f_{\I}$,
\be
f_{\H}(t) = f_{\H}(t_0) + r_{\H} \int_{t_0}^t f_{\S}(t')dt' 
\label{eq:SHIR_DE0_soln} 
\ee
and
\be
f_{\R}(t) = f_{\R}(t_0) + r_{\R} \int_{t_0}^t f_{\I}(t')dt'\ .
\label{eq:SHIR_DE3_soln} 
\ee
Our principle task will therefore be to solve the two coupled equations for $f_{\S}(t)$ and 
$f_{\I}(t)$, the second and third equations in the set (\ref{eq:SHIR_DEs}).

Since the rate of increase of the infected fraction $df_{\I}/dt$ is proportional to 
the product $f_{\I}f_{\S}$, a trivial solution of the model follows from 
an initial condition of no infections, $f_{\I}(t=0) = 0$; this implies  
$f_{\I}(t) = 0$, {\it i.e.} that there are no infections at any later (or earlier) time. 
In this limit all that happens is that the 
initial fully susceptible population $f_{\S}(t=0) = 1$; $f_{\H}(t=0) = 0$ evolves into 
an increasingly hidden population, as a decaying exponential; 

\ba
&f_{\S}(t) &= e^{-r_{\H} t} \ , \cr
\cr
&f_{\H}(t) &= 1 - e^{-r_{\H} t} \ .
\label{eq:SHIR_SHtrivialsoln} 
\ea

In practice the fraction of infected individuals $f_{\I}(t)$ will usually be much smaller than 
the fraction of susceptibles $f_{\S}(t)$. This suggests that we use the no-infection 
limit (\ref{eq:SHIR_SHtrivialsoln}) for $f_{\S}(t)$ as a first approximation in determining 
$f_{\I}(t)$ in the small-$f_{\I}(t)$ limit.
With this substitution, we find that the {\it increase} in $df_{\I}/dt$, the fractional rate of 
new cases per day, will be
\be
\lim_{{\S} >> {\I}} 
df^{(in)}_{\I}/dt    =   + r_{\I} e^{-r_{\H} t} f_{\I} \ .
\label{eq:SHIR_cpd} 
\ee
For completeness we note that the rate of {\it decrease} of the
``active" (still infective) population fraction $f_{\I}(t)$ per day 
due to transitions from ``infected" to ``resolved" status is
given by
\be
df^{(out)}_{\I}/dt    =   - r_{\R} f_{\I} 
\label{eq:SHIR_Ieqout} 
\ee
so the full DE for $f_{\I}(t)$ in this small-$f_{\I}(t)$ limit is
\be
\lim_{{\S} >> {\I}} 
df_{\I}/dt    =   + (r_{\I} e^{-r_{\H} t}  - r_{\R} )f_{\I} \ .
\label{eq:SHIR_IeqsmallfI}
\ee

The formula for the fractional ``new cases per day" (cpd) growth rate in the small-$f_{\I}(t)$ approximation, 
Eq.(\ref{eq:SHIR_cpd}), 
is attractive for several reasons. First, it is easily compared to data, since there is
copious information available from many countries regarding the number of 
new cases per day. (This quantity is simply ${\N}_0$ times the $df^{(in)}_{\I}/dt$ above.) 
Second, since (\ref{eq:SHIR_IeqsmallfI}) is a linear, homogeneous, first-order differential equation 
involving only $f_{\I}(t)$, and the associated integral over time is
tractable, we may solve for $f_{\I}(t)$ analytically in this small-$f_{\I}(t)$ 
approximation. 

In the early and intermediate stages of a pandemic in which the 
infected fraction still dominates the resolved fraction, 
we can ignore the assumed slower transition of infected to resolved by setting $r_{\R} = 0$,
which gives
\be
\lim_{{\S} >> {\I} >> {\R}} 
df_{\I}/dt    =   + r_{\I} e^{-r_{\H} t} f_{\I} \ .
\label{eq:SHIR_Iearlyequn}        
\ee
This relation is especially amenable to comparison with data, as it
relates two observed quantities,
the number of new cases per day (${\N}_0 df_{\I}(t)/dt$) and the cumulative number of cases
(${\N}_0 f_{\I}(t)$), assuming that ${\N}_0 f_{\R}(t)$ can be neglected. 
The SHIR model in this small-infected-fraction limit
predicts that this ratio should be a decaying exponential in time, 
\be
\lim_{{\S} >> {\I} >> {\R}} 
\frac{df_{\I}/dt}{f_{\I}}    =   + r_{\I} e^{-r_{\H} t} \ .
\label{eq:SHIR_fIearly}
\ee

In the following section we will compare the expected time dependence 
of several pandemic models to COVID-19 data, including the SHIR model prediction 
(\ref{eq:SHIR_fIearly}). We will follow this with more detailed SHIR-model calculations 
of the cumulative case numbers ${\N}_0 f_{\I}(t)$ and especially the 
new cases per day (cpd), ${\N}_0 df_{\I}(t)/dt$, and will describe detailed fits 
of these functions to COVID-19 data from a range of countries.

\section{Pandemic Dynamics}

\subsection{Simple models versus COVID-19 data}

Here we will first discuss the time dependence predicted by two familiar dynamic 
models of pandemics, and will show how these models can be compared to COVID-19 data. 
We will find that there are serious difficulties in applying these simple models to COVID-19.

In this discussion we will consider only the cumulative number of cases as a function of time 
(here called ${\P}(t)$), and the rate of appearance of new cases per day, $d{\P}/dt$. 
These numbers are widely reported for COVID-19, and require little interpretation. 
One important point is that the number of cumulative cases to time $t$, ${\P}(t)$, 
is the sum of the infected ${\I}(t)$ {\it and} the resolved ${\R}(t)$ populations 
distinguished in the previous section;
\be
{\P} = {\I} + {\R} \ . 
\label{eq:Pdef}
\ee

As a first, simplest model of the growth of a pandemic, one can assume that 
the rate of change of the number of cases $d{\P}/dt$ is proportional to 
the number of cases ${\P}$,
\be 
\frac{d{\P}}{dt} \equiv \dot {\P} = \lambda {\P} \ .
\label{eq:expDE} 
\ee
The solution of this differential equation is an exponential times the number of cases
at some arbitrary initial time,
\be
{\P}(t) = {\P}(0)\; e^{\lambda t}\ .
\label{eq:expsoln}
\ee
This simple model clearly makes unrealistic assumptions, notably that there 
are infinitely many potential cases. It also neglects spatial dependence 
or other ``local" aspects of the problem, as well as the effect of transitions 
from infected to resolved individuals who are no longer infective.
This exponential model predicts that the ratio of new cases per day to cumulative cases
should be a constant,
\be
\frac{d{\P}/dt}{\P} = \lambda \ .
\label{eq:exp_ratiotest}
\ee

\begin{figure}
\vskip  8mm
\includegraphics[width=0.9\linewidth]{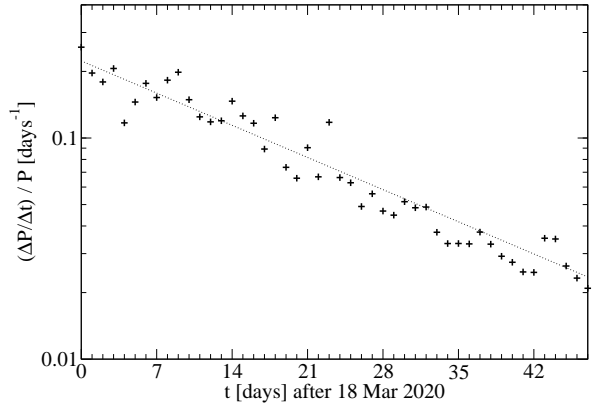}
\caption{The ratio of new cases per day to cumulative cases in UK COVID-19 data.
The exponential model prediction of a constant ratio (\ref{eq:exp_ratiotest}) 
evidently disagrees strongly with this data. The dotted line shows 
an alternative fit to a decaying exponential, which appears as a straight 
line on this log/linear plot.}
\label{fig:ratiotest_UK}
\end{figure}

In Fig.\ref{fig:ratiotest_UK} we compare this expectation to the ratio of 
new cases reported per day ($\Delta {\P} / \Delta t$) to the cumulative number of cases (${\P})$ for
the UK, using data from Ref.\cite{WCU} over the period 18 Mar - 04 May 2020. Evidently this ratio is far from 
constant; it falls by about a factor of 10 over the 47-day period shown. 
 
The simple exponential model can be improved through the introduction of a limit 
in the number of the population that can be infected, ${\P}_{max}$. 
One can assume that the infection rate is initially equivalent to the exponential 
model, but later in the pandemic is suppressed in proportion to the number of 
uninfected individuals remaining, ${\P}_{max}-{\P}$. A simple DE with these properties 
is the ``logistic model," defined by
\be 
\frac{d{\P}}{dt} = \lambda {\P} (1-{\P}/{\P}_{max}) \ .
\label{eq:logisticDE} 
\ee
The solution of this model, which is symmetric about ${\P} = {\P}_{max}/2$, is 
the inverse of an exponential plus a constant, translated by a $t_0$ that is 
determined by the initial conditions;

\be
{\P}(t) = \frac{{\P}_{max}}{\big( 1 + e^{-\lambda (t-t_0)} \big)} \ .
\label{eq:logisticsoln}
\ee

This logistic model can also easily be tested by comparison with 
data on the ratio of new cases per day to cumulative cases. In this case, 
the model predicts that this ratio should be a simple, linear function of ${\P}$, 
\be
\frac{d{\P}/dt}{\P} = \lambda (1-{\P}/{\P}_{max})\ .
\label{eq:logistic_ratiotest}
\ee
So, when data on this ratio (y-axis) is plotted against ${\P}$ (x-axis), this model 
anticipates linear 
${\P}$-dependence, with an x-axis intercept (where $d{\P}/dt = 0$) at ${\P} = {\P}_{max}$. 

\begin{figure}
\vskip  8mm
\includegraphics[width=0.9\linewidth]{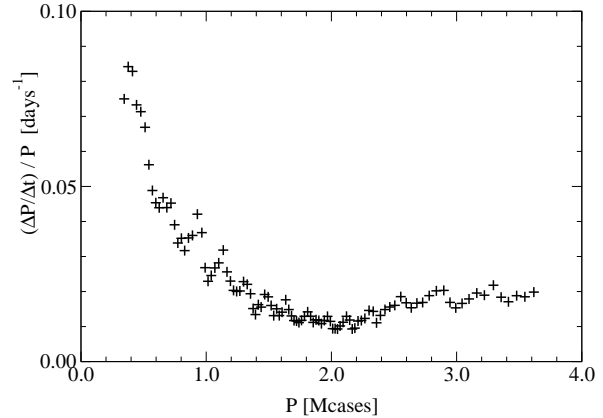}
\caption{The ratio of new cases per day to cumulative cases in USA COVID-19 data (y-axis),
plotted against total cumulative cases (x-axis).}
\label{fig:ratiotest_P}
\end{figure}

In Fig.\ref{fig:ratiotest_P} we show an example of this plot for COVID-19 data, 
using the USA cumulative cases and new cases per day data from Ref.\cite{WCU} 
through 15 July 2020. Although this ratio initially has 
some qualitative resemblance to the logistic model form (\ref{eq:logistic_ratiotest}), 
(see Fig.\ref{fig:ratiotest_P} upper left), 
the more recent data (lower right) is clearly inconsistent with (\ref{eq:logistic_ratiotest}),
and cannot be extrapolated linearly to zero $d{\P}/dt$.

A study of the pandemic time dependence predicted by the logistic model shows another specific
aspect of the model that makes it unrealistic for COVID-19; 
the rate of new cases $d{\P}/dt$ predicted by this model is symmetric about 
the maximum, at the ${\P}(t)$ inflection point $t=t_0$. 
We shall see that in contrast the rate of new cases $d{\P}/dt$ observed in the 
COVID-19 data is rather skewed, and has a faster rise to maximum than the subsequent decline.  

\subsection{The SHIR model versus COVID-19 data}

Here we will test a third prediction for time dependence, which is the form predicted by
the SHIR model of the previous section. There we noted that the infected fraction ${f_{\I}}$ 
of the population is predicted to satisfy the differential equation
\be
\frac{df_{\I}/dt}{f_{\I}}    =   + r_{\I} e^{-r_{\H} t} 
\label{eq:SHIR_IDE}
\ee
in the limits of a small fraction of the population being infected (${f_{\I}} << 1$) 
and the resolved fraction being small relative to the infected fraction 
(${f_{\R}} << f_{\I}$). The existing COVID-19 data suggests that these 
are both reasonable approximations in the early and intermediate stages of 
the COVID-19 pandemic.

The expectation (\ref{eq:SHIR_IDE}) can immediately be compared to data, since 
$(df_{\I}/dt)/f_{\I}$ is just the ratio shown on the y-axis of Fig.\ref{fig:ratiotest_UK}. 
The dotted line in the figure shows a decaying exponential fit of exactly the form
(\ref{eq:SHIR_IDE}), which evidently gives a very good description of the data. 
The parameter values found in this fit are $r_{\I} = 0.233$~[days$^{-1}$] and 
$\tau \equiv 1/r_{\H} = 20.9$~[days].

Motivated by the good agreement between the SHIR model prediction of a decaying exponential
in time for $(df_{\I}/dt)/f_{\I}$ and the COVID-19 data evident in Fig.\ref{fig:ratiotest_UK},
we will subsequently discuss SHIR model predictions in more detail, and will 
show fits of this model to COVID-19 data from many European countries.
This will follow a short discussion of some general aspects of mathematically similar models,
which will be used in the detailed discussion of SHIR results.

\subsection{General results for a class of pandemic models}

The results discussed above motivate the consideration of generalizations of the simplest 
``pure exponential" pandemic model, in which the constant $\lambda$ of that model
(\ref{eq:exp_ratiotest})
is replaced by an explicit function of time,
\be
\frac{d{\P}/dt}{\P} \equiv \lambda(t) \ . 
\label{eq:defnlambdaoft}
\ee

For a given function $\lambda(t)$, the defining differential equation 
(\ref{eq:defnlambdaoft}) gives the predicted growth of case numbers as 
\be
{\P}(t) = {\P}(0)\; e^{\int_0^t \lambda(t') dt'}.
\label{eq:Poftgensoln}
\ee
Formally this may be regarded as exponential growth in a new (nonlinear) time 
variable $u$, defined by $du = \lambda(t) dt$.

Since this is a model of the {\it growth} of a pandemic, the cumulative number of cases 
(\ref{eq:Poftgensoln}) can only increase, or 
become flat when the pandemic has been completely halted, 
which occurs if and when we reach a time $t_f$ for which
$\lambda(t_f)=0$. Accordingly we constrain $\lambda(t)$ to be positive 
(pandemic ongoing, $t < t_f$) or zero ($t = t_f$, pandemic halted).

The inflection point of ${\P}(t)$ is of considerable importance, since this is the 
time of peak pandemic growth.  The time $t_{infl}$ of the inflection point 
is specified by the rate of change of case numbers $d{\P}/dt$ reaching a maximum; 
this occurs when the second derivative $d^2{\P}/dt^2$ is zero,
\be
\frac{d^2{\P}(t)}{dt^2} \Big{|}_{t=t_{infl}} = 0\ .
\label{eq:deftinfl}
\ee
On substituting the general solution ${\P}(t)$ from (\ref{eq:Poftgensoln})
into this constraint, we find an equivalent definition of the inflection point 
in terms of the proportional growth function $\lambda(t)$,
\be
\Bigg\{\frac{d\lambda(t)}{dt} + \lambda(t)^2\Bigg\}\Bigg{|}_{t=t_{infl}}  = 0\ .  
\label{eq:deftinfllambda}
\ee

\section{Predictions of the SHIR model}

\subsection{Cumulative cases ${\P}(t)$ and cases per day (cpd) $d{\P}/dt(t)$ }

In Sec.I.B. we introduced the ``SHIR" model as a generalization of the SIR model, 
with the effects of pandemic reduction measures incorporated through 
a ``hidden" population (H) that migrates from the initial susceptible population 
(S) with a rate constant $r_{\H}$. In the limit of a small infected population fraction
($f_{\I} << 1$) relative to the remaining fraction of susceptibles ($f_{\S}$), and assuming
that the resolved fraction $f_{\R}$ can be neglected, we showed that the infected fraction 
satisfies a differential equation of the form
\be
\lim_{{\S} >> {\I} >> {\R}} 
\frac{df_{\I}/dt}{f_{\I}} = \lambda(t)  
\label{eq:SHIR_soln1}
\ee
where the SHIR growth rate function is a decaying exponential in time,
\be
\lambda(t)  =   + r_{\I} e^{-r_{\H} t}\ .
\label{eq:SHIR_soln2}        
\ee
The rate coefficients $r_{\H}$ and $r_{\I}$ can be interpreted respectively as a characteristic 
``social response time"  
$\tau = 1/r_{\H}$ for the implementation of pandemic reduction measures,
and an initial condition (at some $t=0$) on the growth rate function $\lambda(t)$; 
\be
\lambda(t) = \lambda(0)\, e^{-t/\tau}\ .
\label{eq:SHIR_lambdaoft}
\ee 

If we multiply the infected fraction $f_{\I}$ by the total population ${\N}_0$, 
\be
{\P}(t) = {\N}_0 \cdot f_{\I} 
\label{eq:def_Poft}
\ee
and similarly scale the fractional new cases per day, 
\be
d{\P}/dt = {\N}_0 \cdot df_{\I}/dt\; ,  
\label{eq:def_dPdt}
\ee
we can compare the results for the cumulative cases ${\P}(t)$ and new cases per day 
$d{\P}/dt$ directly to the data as normally reported.

Given the form (\ref{eq:SHIR_lambdaoft}), on solving (\ref{eq:SHIR_soln1}) we find that the 
predicted number of cases as a function of time in the SHIR model is a nested exponential,
\be
{\P}(t) = {\P}_{max} / \exp\big( \lambda(0) \tau\, e^{ -t/\tau } \big) .
\label{eq:SHIR_Poftpre}
\ee 
This may be simplified by replacing the $t=0$ initial condition  
$\lambda(0)$ by an equivalent time shift
\be
t_0 = \tau \ln (\lambda(0)\tau )\ ,
\label{eq:SHIR_t0}
\ee
which gives a simple three-parameter form for the general solution of the 
cumulative number of cases in the SHIR model,
\be
{\P}(t) = {\P}_{max} / \exp\big( e^{ -(t-t_0)/\tau } \big) .
\label{eq:SHIR_Poft}
\ee
This function increases monotonically from ${\P}=0$ at $t=-\infty$ to ${\P}_{max}$ at 
$t=+\infty$, and has a single inflection point at $t = t_0$.
This is an example of a Gompertz \cite{Gom1832} curve, $y = ab^{q^x}$\cite{Wei}; 
these are often encountered in problems in which a proportional rate of growth 
$\dot f / f$ is driven by an exponential. 

The rate of appearance of new cases per day (cpd) in the model, $d{\P}/dt$, 
can be evaluated directly from ${\P}(t)$ above, and is 
\be
\frac{d{\P}}{dt} = 
\frac{{\P}_{max}}{\tau} \cdot 
e^{-(t-t_0) /\tau} / \exp\big( e^{ -(t-t_0)/\tau } \big).
\label{eq:SHIR_dPdt}
\ee
This rate of growth is zero at $t = -\infty$, increases to a maximum value at the 
${\P}(t)$ inflection point $t_0$, and then decreases with time, approaching zero exponentially 
as $t\to +\infty$. $d{\P}/dt$ integrates to a finite number of cases ${\P}_{max}$ as $t\to +\infty$. 
The maximum number of cases per day, at the time $t_0$ of the 
${\P}(t)$ inflection point, is given by 
\be
\frac{d{\P}}{dt}\bigg{|}_{t=t_{infl}=t_0} =
\frac{{\P}_{max}}{e\tau }\ .
\label{eq:SHIR_dPdtmax}
\ee

The three parameters that specify a solution of the SHIR model, 
${\P}_{max}$, $\tau$ and $t_0$, 
all have simple interpretations; they are respectively 
the height, width scale, and time shift 
of the pandemic profile ${\P}(t)$ (or of the rate of new cases, $d{\P}/dt$). 
${\P}_{max}$ is by definition the maximum value of ${\P}(t)$, 
and gives the total number of cases, which is ${\P}(t)$ at $t=+\infty$.
The social response time $\tau$ sets the scale for {\it all} 
time-interval features of ${\P}(t)$ and $d{\P}/dt$, 
since changing $t_0$ only translates these functions.
The time shift $t_0$ is the time of the inflection point of ${\P}(t)$, 
\be
t_{infl} = t_0 = \tau \ln (\lambda(0) \tau ) .
\label{eq:SHIR_tinfl}
\ee
This inflection point $t_0$ may be considered the peak 
of the local pandemic, since the number of new cases per day estimated 
by the fit is then a maximum.

\begin{figure}[h]
\vskip  5mm 
\includegraphics[width=0.9\linewidth]{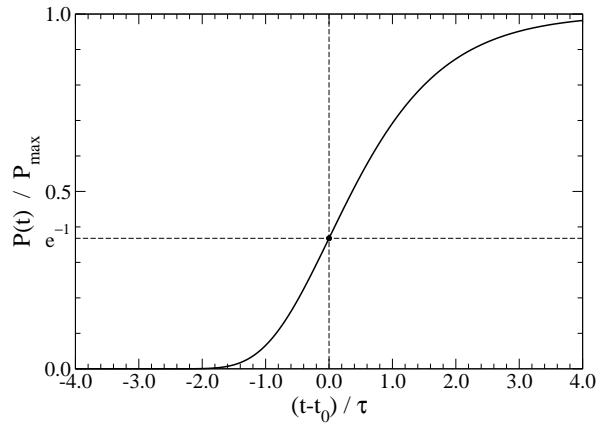}
\vskip  -2mm 
\caption{A ``universal" plot of the cumulative case numbers ${\P}(t)$ predicted 
by the SHIR model (\ref{eq:SHIR_Poft}). To the extent that the model 
is accurate, all cumulative case number data should follow this curve.}  
\label{fig:Poft_univ}
\end{figure}

For discussion purposes a convenient choice for $t_0$
in (\ref{eq:SHIR_Poft}) and (\ref{eq:SHIR_dPdt}) is $t_0 = 0$, which places the inflection point 
of ${\P}(t)$ at $t=0$. Plotting ${\P}(t)$ or $d{\P}/dt$ against $t$ in units of $\tau$, 
normalized to their maximum values, gives a parameter-independent 
``universal profile" for each of these functions.  
The universal form for ${\P}(t)$, scaled by its maximum value ${\P}_{max}$,
is shown in Fig.\ref{fig:Poft_univ}. The inflection point of ${\P}(t)$ at $t=0$ is indicated,
where both ${\P}(0)/{\P}_{max}$ and its slope 
(in the dimensionless time variable $t/\tau$) equal $e^{-1}$. 
This last property implies that the total number of cases ${\P}_{max}$ and the cumulative 
number of cases at the time of the ${\P}(t)$ inflection point (pandemic maximum), ${\P}(t_0)$, are related by
\be
{\P}_{max} = e\cdot {\P}(t_0) \ .
\label{eq:Pmax_estm}
\ee
This provides a simple estimate of the total expected cases ${\P}_{max}$ from the number 
observed up to the pandemic peak, ${\P}(t_0)$, which may already be known. 

Some additional times of interest are those 
at which ${\P}(t)$ has reached $10\%$, $50\%$, $90\%$ and $99\%$ 
of the total cumulative cases; these are respectively 
$(t-t_0)/\tau = -0.8340, +0.3665, +2.250,$ and $+4.600$. 

\begin{figure}[h]
\vskip  1mm 
\includegraphics[width=0.9\linewidth]{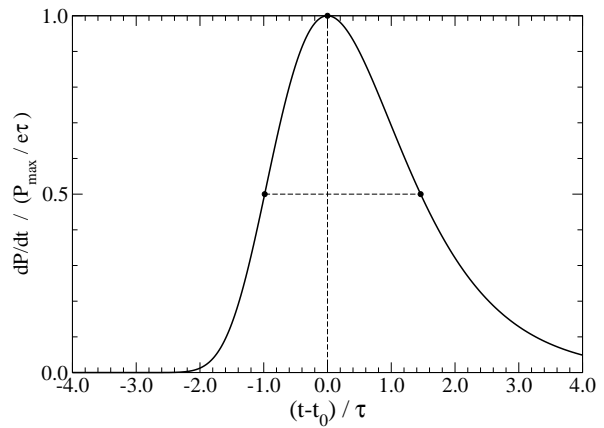}
\vskip  -2mm 
\caption{The corresponding universal plot of new cases per day (cpd), 
$d{\P}/dt$, predicted by the SHIR model (\ref{eq:SHIR_dPdt}). 
The cpd data should follow this curve if the SHIR model is accurate.}  
\label{fig:dPdt_univ}
\end{figure}

The universal curve for $d{\P}/dt$, the new cases per day, is shown in 
Fig.\ref{fig:dPdt_univ}, scaled by its maximum value of ${\P}_{max} / e\tau $.
The mean and variance of $t$ for this distribution are $<t> = \gamma \tau$
($\gamma$ is Euler's constant) and $\sigma = (\pi / \sqrt{6})\tau$. 
The skewness \cite{def_skewness} of $d{\P}/dt$ is 
$12\sqrt{6}\,\zeta(3)/\pi^3 \approx +1.140$, 
a positive value indicating a more heavily weighted RHS.
Several additional interesting features of $d{\P}/dt$ are the FWHM, $2.446 \tau$ 
(indicated in Fig.\ref{fig:dPdt_univ}), the rise time of $d{\P}/dt$ 
from half-max to peak, which is $0.985\tau$; and the much slower decline of 
$d{\P}/dt$ from peak to half-max, which requires $1.461\tau$. 

The asymptotic approach of ${\P}(t)$ to ${\P}_{max}$ cases at large times is exponential,
\be
\lim_{t\to\infty} {\P}(t)/{\P}_{max} = 1 - e^{ -(t-t_0)/\tau },
\label{eq:SHIR_Pofttoinfty}
\ee 
as is the (decreasing) rate of growth of new cases at large times,
\be
\lim_{t\to\infty} \frac{d{\P}}{dt} =
({\P}_{max} / \tau ) \, e^{-(t-t_0)/\tau}\ . 
\label{eq:SHIR_dPdtatlarget}
\ee
We note in passing that these results suggest an improved estimate of the asymptotic total number 
of cases ${\P}_{max}$ which has accelerated convergence,
\be
\lim_{t\to\infty} 
\Big( {\P}(t) + \tau \frac{d{\P}}{dt} \Big) = {\P}_{max} + O(e^{-2(t-t_0)/\tau}) ,
\label{eq:SHIR_Pmaxestimator}
\ee
although this improved behavior would likely be masked in practice by the background of sporadic cases.
 
\section{Application of the SHIR model to COVID-19 Data}

\subsection{Goals of the Fitting Exercise}

In this section we will show results from fitting the 
predicted early pandemic time dependence from the SHIR model 
as derived in Sec.IIIA to data for the confirmed COVID-19 case numbers 
from many different countries. 
We will primarily consider fits of the new case numbers per day (cpd)
to the SHIR model prediction, Eq.(\ref{eq:SHIR_dPdt}). 
As an initial example we will show results for Austria, including fits to both 
the cpd and the cumulative case numbers. Following this we will give the
results of SHIR model fits to the cpds reported by a representative set of
Western countries, and will compare and contrast these results.
We will also discuss procedures for fitting more complicated datasets that show evidence for 
multiple infection sites, and will show examples of such fits.  
Finally we will show how datasets from multiple countries can be combined, and will use
this procedure to test the predicted SHIR model profile for new cases per day, displayed as a
universal (parameter-free) curve.

\subsection{SHIR model single country fits: Europe}

As a first application we will fit the three-parameter SHIR model 
to the data on the number of new daily COVID-19 cases per day in Austria,
as reported on the Worldometer website, Ref.\cite{WCU}. 
Austria was chosen because it represents one of 
the earliest COVID-19 outbreaks in Europe, so their data 
allows us to follow both the earlier {\it and} later stages 
of the pandemic. For this fit we will use the entire 
Austrian cpd dataset from the initial website date of 15 Feb 2020 
through 06 May 2020, comprising a nominal 82 data points. 
(For most countries we consider, including Austria, there are actually many ``null'' or zero entries 
in the early dates.)  

As a reminder, in the small fraction of infections limit assumed here one may solve the model 
for the cumulative case numbers of infected people ${\P}(t)$ and hence the rate of new cases 
per day $d{\P}/dt$ in closed form, given by Eqs.(\ref{eq:SHIR_Poft},\ref{eq:SHIR_dPdt}). 
The three free parameters in these functions are ${\P}_{max}$ (total number of cases), 
$\tau$ (social response time), and 
$t_0$ (the ${\P}(t)$ inflection point; time of the maximum cases per day). 
Our common $t=0$ reference time in all these numerical fits and discussions is taken to be 
18 Mar 2020. The fits are all simple, unweighted least-squares without data errors.

\begin{figure}[ht]
\vskip  5mm
\includegraphics[width=0.9\linewidth]{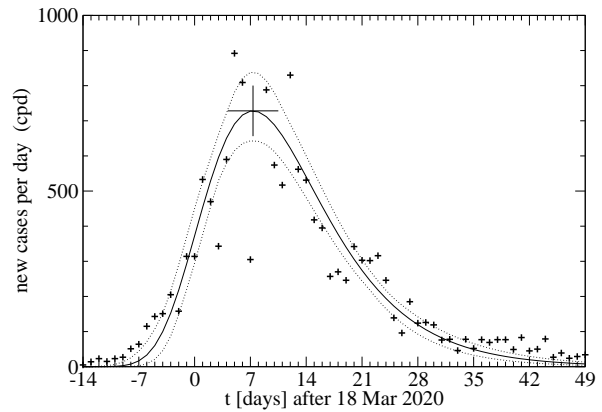}
\caption{New COVID-19 cases per day reported for Austria over the period 15 Feb - 06 May 2020, 
with a SHIR model fit (solid line). 
The maximum new case rate and associated time (the ${\P}(t)$ inflection point, $t_0$) in the model 
are indicated by a single large point. 
Estimated 95\% CL limits are shown as dotted lines.} 
\label{fig:Austria_cpd_SHIRfit}
\end{figure}
The result of this fit for Austria is shown in Fig.\ref{fig:Austria_cpd_SHIRfit}.
This is a surprisingly good fit to the full range of the cpd data, including the rapid 
initial rise after the early cases, the peak region, and the subsequent slower decline; 
all are reasonably well described by the model 
\cite{AusNorWCUpts}.

The SHIR model parameters resulting from this fit to the cpd data from Austria 
are given below.

\ba
&{\P}_{max}       &=  14.9(8)k\  [\hbox{\rm cases}] \cr
&\tau             &=   7.54(50)\  [\hbox{\rm days}] \cr
&t_0              &=   7.32(49)\  [\hbox{\rm days}] \ .
\label{eq:Austria_cpd_SHIRfit}
\ea
These and other fitted parameters are typically quoted to three-place accuracy in this paper,
to allow numerical tests such as debugging of numerical routines and checking analytic results.  

There is evidently reasonable agreement between the SHIR model 
and the data with these parameters for the number of new cases per day, 
although there is considerable scatter in the data. The numerical value 
of the peak predicted cpd, ${\P}_{max}/e\tau$, and the time of the peak, 
$t_{infl} = t_0$ are indicated by a single large point
in Fig.\ref{fig:Austria_cpd_SHIRfit}. These values are respectively 
728~[cases/day] and 7.32~[days] (measured from $t=0$ on 18 Mar 2020), 
giving 25 Mar 2020 as the peak date for new cases per day in Austria implied by this fit. 

The fitted value of the Austrian ``social response time," $\tau \approx 7.5$~[days], 
is of special note. This is the smallest value (hence the fastest social response) 
we have found in our fits to the COVID-19 pandemic numbers from a large set of 15 representative 
Central and Western European countries (see Table~\ref{table:SHIR_EUfitbycountry}).
This suggests that Austria deployed an especially rapid and effective response 
to this pandemic, which presumably had a strongly limiting effect on the total number of 
cases. 

A minor discrepancy between the model and data is evident at the early times
in Fig.\ref{fig:Austria_cpd_SHIRfit} that are characterized by small case numbers. 
(See times before about $t = -4$~[days] in the figure.) 
The number of early cases is clearly underpredicted by the fit. 
This may represent their growth from a small number of initial cases, 
closely monitored and hence developing rather slowly, 
which were evident before the primary pandemic contribution became dominant. 
We will note a similar, more prominent effect in the early data from 
Denmark and Norway at about the same date.

An alternative approach would be to fit the cumulative number of reported cases
as a function of time, which is typically reported together with the new cases
per day. Since one is the derivative of the other, these should 
give similar parameter values. Of course the results will not be identical, 
because the model is not exact, and the cumulative cases data will give 
higher fitting weight to the later stages of the pandemic. 

The choice of which dataset to fit, cumulative cases or new cases per day, 
may be suggested by the subject of greatest interest; 
the peak of the pandemic is best determined by fitting the new cases per day, 
but the final number of cases will be best determined if constrained by 
the large total case numbers reported in the later stages of the pandemic.  

\begin{figure}[h]
\vskip -1mm
\includegraphics[width=0.9\linewidth]{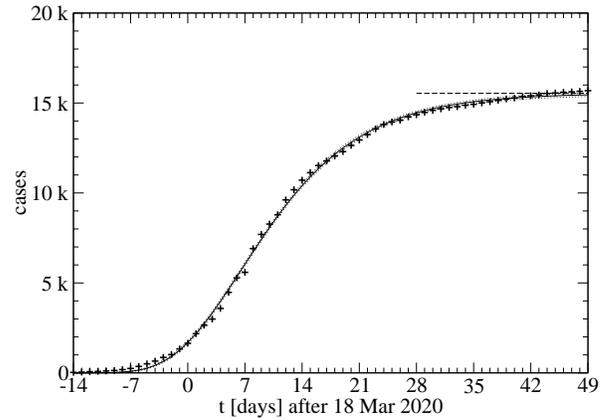}
\vskip -2mm
\caption{A fit of the SHIR model to Austrian cumulative COVID-19 cases,
showing the data and the corresponding model results for 
${\P}(t)$ (solid line).
Estimated 95\% CL limits are included as dotted lines, but are very close to the central curve. 
The asymptotic number of cases predicted by this fit, $15.54k$, is shown as 
a horizontal dashed line.}
\label{fig:Austria_cas_SHIRfit}
\end{figure}

Here as an example of the second fit we give the SHIR model results that follow from 
fitting the cumulative case numbers from Austria. We again use a set of 82 points 
of Austrian data~\cite{WCU} from the initial reporting date of 15 Feb 2020 through the 
fitting date, 06 May 2020. 
The SHIR model fit to the Austrian cumulative cases data is shown in 
Fig.\ref{fig:Austria_cas_SHIRfit}. 
The parameters resulting from this cumulative cases fit are
\ba
&{\P}_{max}       &=  15.54(5)k\  [\hbox{\rm cases}] \cr
&\tau             &=   8.18(11)\  [\hbox{\rm days}] \cr
&t_0              &=   6.53(7) \  [\hbox{\rm days}] 
\label{eq:Austria_cas_SHIRfit}
\ea
Evidently the SHIR model also provides a reasonably accurate description of the cumulative 
cases for Austria.

The parameter values resulting from the two fits, to cpd data (\ref{eq:Austria_cpd_SHIRfit}) 
and cumulative cases data (\ref{eq:Austria_cas_SHIRfit}), are 
quite similar \cite{cas_fit}. 
The total case numbers ${\P}_{max}$ differ from their mean by $\pm 2\% $, 
the fitted social response times $\tau$ differ from their mean by $\pm 4\% $, 
and the fitted times of the Austrian inflection point (maximum rate of new cases) 
in the two fits differ by just 0.8~[days].  

Next we will follow the same procedure described in the first Austrian fit above, 
and will fit the SHIR model to the cpd data for each of a representative 
set of Central and Western European countries. For each fit, as discussed for Austria, 
we have again used 82 pts of daily new case numbers per day  
\cite{WCU} for the period 15 Feb - 06 May 2020 as the only input data. 
This data fitted to the three-parameter SHIR model form for $d{\P}(t)/dt$, 
(\ref{eq:SHIR_dPdt}). This fixes the three model parameters ${\P}_{max}$ 
(with units of [cases]), $\tau$~[days], and the inflection point 
$t_{infl} = t_0$~[days] (measured from $t=0$ on 18 Mar 2020) separately 
for each country. 

The fitted parameter values for the entire set of 15 Central and Western European 
countries considered here are given in Table.\ref{table:SHIR_EUfitbycountry}, 
together with the calendar date of the inflection point $t_0$, and the fitted peak 
number of new cases per day, which occurs at $t_0$.

\begin{table}[h]
\begin{center}
\begin{tabular}{|r|c|c||c|c|c|}
\hline
\ & & & & &
\\
Country
& ${\P}_{max}$ 
& $\tau$ 
& $t_0$ 
& peak date
& max cpd 
\\
\hline
\ & & & & &
\\
\
Austria
& 14.9\k
& 7.54 
& 7.32
& 25 Mar
& \phantom{0}728
\\
\
Iceland
& 1.80\k
& 8.69 
& 7.55
& 26 Mar
& \phantom{00}76
\\
\
Switzerland
& 30.2\k 
& 10.1 
& 8.02
& 26 Mar
& 1100
\\
\
Norway$^*$ 
& 6.95\k
& 10.1 
& 8.71
& 27 Mar
& \phantom{0}252
\\
\
France
& 175\k 
& 10.5 
& 16.3
& 03 Apr
& 6100
\\
\
Germany
& 172\k
& 11.3 
& 12.0 
& 30 Mar
& 5590
\\
\
Ireland
& 24.0\k
& 11.8 
& 26.2 
& 13 Apr
& \phantom{0}750
\\
\
Portugal
& 28.6\k 
& 13.2
& 17.6 
& 05 Apr
& \phantom{0}599
\\
\
Belgium
& 57.4\k
& 14.2 
& 19.8
& 07 Apr
& 1491
\\
\
Spain
& 273\k
& 14.2 
& 14.4
& 01 Apr
& 7010
\\
\
Denmark$^*$
& 10.2\k
& 14.3 
& 19.6
& 07 Apr
& \phantom{0}262
\\
\
Netherlands
& 46.9\k 
& 15.2 
& 18.0
& 05 Apr
& 1140
\\
\
Italy
& 232\k
& 16.1 
& 9.72
& 28 Mar
& 5310
\\
\
UK
& 302\k
& 20.4 
& 30.9 
& 18 Apr
& 5450
\\
\
Sweden
& 40.9\k
& 25.7 
& 33.8
& 21 Apr
& \phantom{0}586
\\
\hline
\end{tabular}
\vskip 2mm
\caption{The result of fitting the SHIR model to the data on COVID-19 
new cases per day from 15 representative Central and Western European countries. 
The first two model parameters shown are the predicted total cumulative COVID-19 cases
${\P}_{max}$ and the social response time $\tau$. The entries are ordered by $\tau$.
The final three columns give the fitted inflection point $t_0$ (the pandemic peak) 
[days after 18 Mar 2020] and date, and the fitted maximum cpd (at $t_0$).
\hskip 5mm   $^*$See discussion.}  
\label{table:SHIR_EUfitbycountry}
\end{center}
\end{table}

\subsection{Implications of European Single Country Fits}

\begin{figure}[h]
\includegraphics[width=0.9\linewidth]{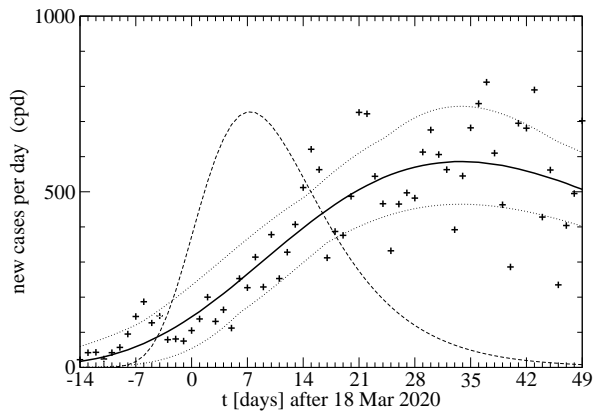}
\caption{The SHIR model fit to the Swedish COVID-19 new cases per day data 
over the period 15 Feb - 06 May 2020. 
Estimated 95\% CL limits are shown as dotted lines.
The corresponding fit to the Austrian cpd data of
Fig.\ref{fig:Austria_cpd_SHIRfit} is shown for comparison as a dashed line.}
\label{fig:Austria_vs_Sweden_cpd}
\end{figure}

Perhaps the most remarkable aspect of the SHIR model fits to 15 Central and Western European
countries in Table.\ref{table:SHIR_EUfitbycountry} is the wide range of values found for the
``social response time" $\tau$, and how it appears to correlate with national 
coronavirus policy. As $\tau$ sets the width of the pandemic peak, it is a measure of how long
a country will have to endure the pandemic. (Recall for example that the FWHM of the 
new cases per day, $d{\P}/dt$, is about $2.4\cdot \tau$.)
The extreme cases are Austria ($\tau \approx 7.5$~days) and Sweden ($\tau \approx 26$~days), 
over 3 times longer than Austria. The Swedish cpd data and fit are shown in 
Fig.\ref{fig:Austria_vs_Sweden_cpd}.

The very slow Swedish response time is apparently the result of a rather controversial 
COVID-19 policy promoted by the Swedish State Epidemiologist, A.~Tegnell~\cite{Sweden_policy}, 
which did not impose the strict school closings, border controls, and restrictions on 
large gatherings that characterized the policies of other Scandinavian countries. 
Tegnell instead advocated a more open society during the pandemic, to promote 
herd immunity and a quick economic recovery. Although the Swedish policy was evidently 
successful in broadening the curve, their {\it per capita} number of 
cases is over twice that of Denmark and Norway, and the Swedish fatalities 
{\it per capita} are larger by a similar amount. This argues that an aggressive
national policy of minimizing the social response time $\tau$, for example through 
contact tracing, extensive testing, rapid medical response and isolation of positive cases
through quarantine, may be the most effective in reducing the total case and fatality numbers. 
It may also prove easier to sustain a strict policy over a short period of time than a more lax one 
over a longer period.

The UK and Dutch social response times $\tau$ are also quite large, and can also be correlated with 
their early national policies that promoted ``flattening the curve" and developing herd immunity.
To quote the Dutch Prime Minister regarding COVID-19 policy in a 
national address~\cite{Netherlands_policy}, (in translation) 
``... {\it we can slow down the spread of the virus while at the same time building group 
immunity in a controlled way.}" In a national address~\cite{UK_policy} the UK Prime Minister stated that 
``{\it It is vital to slow the spread of the disease. That is the way we reduce the number 
of people needing hospital treatment at any one time.}" 

Of course the argument that slowing the spread of the pandemic is advantageous 
because it ``flattens the curve" is specious. This tacitly assumes that the total number of 
cases ${\P}_{max}$ is constant, so a broader rate curve must have a lower mean. 
In practice there is instead evidence that a longer $\tau$ may result in more infections 
{\it per capita}, as seen in comparing Sweden with Denmark and Norway. 
If so, flattening the curve by slowing the spread of the pandemic unfortunately 
increases the total number of cases and fatalities. This complicated and perhaps 
counterintuitive issue certainly merits careful future investigation.

\subsection{SHIR model single country fits: North America}

We have also carried out fits of the SHIR model to COVID-19 data for three North American countries,
Canada, Mexico and the USA. These three countries were treated as a separate exercise because of concerns 
that European and North American policies differed considerably, which might lead to very different fit 
parameters and pandemic curves. This was indeed found to be the case. 

Most of the fits of the SHIR model to European countries described above were carried out in
May 2020, using 82-point Worldometer coronavirus datasets \cite{WCU} covering the period 
15 Feb - 06 May 2020. Initially we used the same procedures and datasets for the three 
North American countries. This appeared satisfactory for Canada and the USA, however the data 
for Mexico showed that this country was still in the early stages of the pandemic, and had a 
very long social response time of about 63 days, with a predicted pandemic peak in late July. 
For this reason we decided to carry out a fit to Mexico incorporating later data, specifically 
a 151-point Worldometer cpd dataset for the period 15 Feb - 14 July 2020, and have fitted 
the corresponding data for Canada as well. 

Recent developments in the USA have shown a dramatic departure from a single peak model, 
which suggests that the USA should be treated as a superposition of 
several more localized outbreaks. Accordingly, here we quote SHIR model fit results 
for New York State alone, as a representative component of the USA with an effective social 
response. The cpd data in this case is available from 12 Mar 2020, so we fitted a dataset covering 
the period 12 Mar - 14 July 2020, comprising 125 data points.

The results of these fits are given in Table.\ref{table:SHIR_NAfitbycountry}. 
The social response times $\{\tau\} $ for North America extend from near to well 
beyond the upper end of the European range; the values for New York State and 
Canada are comparable to the highest values observed in Europe, 
near the 16~[days] and 26~[days] found for Italy and Sweden respectively. 
This suggests comparably extended periods of pandemic in these North American countries.
The $\tau$ value of about 65~[days] found in the fit to Mexico is remarkable, 
much longer than that of any other country considered in this study. 
(A fit to Brazilian data, not included here, finds a similar value for $\tau$.)
This SHIR model fit for Mexico (to the 15 Feb - 14 July 2020 cpd data) predicts that their 
pandemic peak will occur on 31 July 2020, and anticipates just over a million total cases.     

\begin{table}[h] 
\begin{center}
\begin{tabular}{|c|c|c||c|c|c|}
\hline
\ & & & & &
\\
Region
& ${\P}_{max}$ 
& $\tau$ 
& $t_0$ 
& peak date
& max cpd 
\\
\hline
\ & & & & &
\\
\
NY State 
& 404\k
& 15.7
& 20.6 
& 08 Apr
& 9480
\\
\
Canada 
& 112\k
& 24.6 
& 35.4 
& 22 Apr
& 1680
\\
\
Mexico 
& 1.11\M
& 65.2
& 135.2 
& 31 Jul
& 6280
\\
\hline
\end{tabular}
\vskip 2mm
\caption{The result of fitting the SHIR model to data on COVID-19 new 
cases per day in North American countries, ordered by social response 
time $\tau$. The entries are as in Table.\ref{table:SHIR_EUfitbycountry}.}  
\label{table:SHIR_NAfitbycountry}
\end{center}
\vskip -1cm
\end{table}

\subsection{Fits to countries with multiple events}

Of course each of the national pandemic datasets we are studying is in some sense the sum of
multiple events, displaced in time and encountering special local conditions. 
In the USA for example the early data is dominated by events on the West Coast, 
but these were superseded by the major outbreak involving New York City, and then by a smaller
event in New Jersey, and a larger number of local events in the Midwest and South. Fitting the 
data from many local outbreaks as a single event is clearly problematic. 
A series of local events separated in time but still overlapping significantly, 
if fitted as a single outbreak, could give a misleadingly large $\tau$. 
Separating the individual events will require accurate local data, and a more detailed investigation.  

Despite this concern, the SHIR model does nonetheless allow a reasonably accurate fit to most of
the national datasets we have considered. Of the 15 European countries we have discussed here, 
just two showed clear evidence of a more complicated pandemic history that could not be fitted 
as a single-peak event. These were Denmark and Norway. In both countries there was evidence of an early, 
short-lived outbreak, which led to about 800 cases of COVID-19 in each country, both
during the period of 10-13 Mar 2020. In both countries this early outbreak was rapidly suppressed, 
presumably through contact tracing and quarantine.

\begin{figure}[h]
\includegraphics[width=0.9\linewidth]{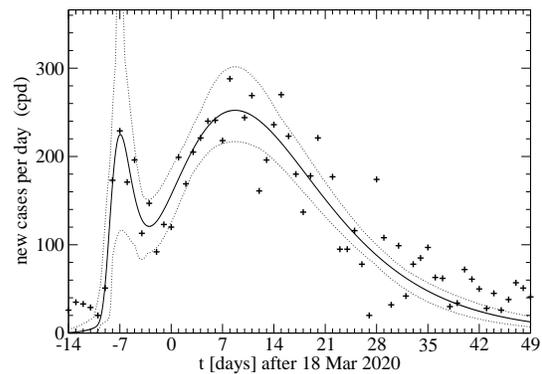}
\caption{A two-peak fit to the COVID-19 cpd data of Norway, using the 
form (\ref{eq:SHIR_dPdt}) with two independent parameter sets.
Estimated 95\% CL limits are shown as dotted lines.} 
\label{fig:Norway_cpd_twopeaks}
\end{figure}

We found that the cpd data from Denmark and Norway could be described 
by using Eq.(\ref{eq:SHIR_dPdt}) to represent two independent outbreaks, a short, early one,
with a very short $\tau \approx 1.2$~[days], and a dominant second outbreak, 
with the more typical COVID-19 parameters listed in Table.\ref{table:SHIR_EUfitbycountry}.
The resulting two-peak fit for Norway is shown in Fig.\ref{fig:Norway_cpd_twopeaks} 
\cite{AusNorWCUpts}.

The fitted parameters for the early Norwegian peak are 
${\P}_{max} = 764$~[cases], $\tau = 1.44$~[days], and $t_0 = -7.06$~[days].
The very similar early peak in Denmark gave the parameters
${\P}_{max} = 778$~[cases], $\tau = 1.22$~[days], and $t_0 = -7.21$~[days].
The total cases predicted for Denmark and for Norway are simply the sum of the 
early and later peaks, thus $11.01k$ and $7.72k$ cases respectively. 

\subsection{SHIR model fits to multiple countries}

The idea of simultaneously fitting multiple countries to the SHIR model seems appealing, 
particularly as this model gives a good description of the case numbers from many different 
European countries. Unfortunately there are complications with this procedure; 
individual countries vary considerably in the fitted values of the total cases ${\P}_{max}$, 
the characteristic social response time $\tau$, and the local time of the pandemic peak, $t_0$. 
These differences suggest a scaling approach, in which we determine these parameters separately 
for each country in a SHIR model fit, and then translate (by $t_0$) and scale 
(by ${\P}_{max}$ and $\tau$) the datasets for each country separately. If the model and the data 
are both accurate, these translated and scaled datasets should provide points that fall 
on the universal curves shown in Figs.\ref{fig:Poft_univ} and \ref{fig:dPdt_univ}. 
Of course there are indications that the COVID-19 data itself is occasionally problematic, 
with considerable scatter, single-bin spikes, and in some countries a modulation with 
a 7-day period that may reflect a weekly reporting regimen. Nonetheless it is of interest 
to generate this translated and scaled scatterplot, to see whether a common pandemic 
profile is indeed evident in the European data. 

\begin{figure}[h]
\includegraphics[width=0.9\linewidth]{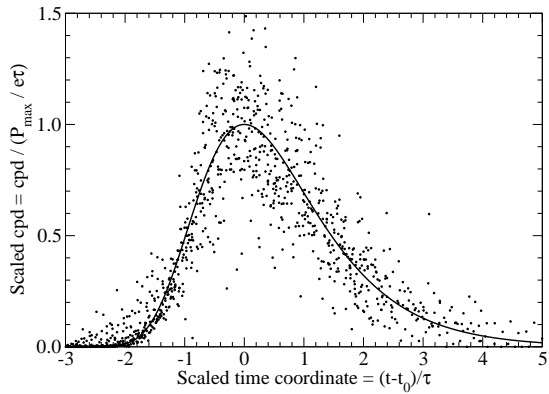}
\vskip -2mm
\caption{A universal scatterplot of 1162 cpd data points from the 15 
European countries of Table.\ref{table:SHIR_EUfitbycountry}, translated and scaled to fit on the 
universal plot for $d{\P}/dt$, Fig.\ref{fig:dPdt_univ}. The (parameter-free) SHIR model prediction 
for $d{\P}/dt$, Eq.(\ref{eq:dPdt_univ_infig}), is shown as a solid line in this figure.} 
\label{fig:AllEurope_cpd}
\end{figure}

This scatterplot is shown in Fig.\ref{fig:AllEurope_cpd} for the 82-point cpd datasets 
of all 15 European countries in Table.\ref{table:SHIR_EUfitbycountry}. Each national 
dataset was translated and scaled separately by its particular values of $t_0$, 
${\P}_{max}$, and $\tau$ (as given in the table) before being added to this figure. 
The data from Denmark and Norway before 20 Mar 2020 was excluded from the plot, 
to suppress contributions from their initial early outbreaks. 

The scaled, centered universal curve for $d{\P}/dt$ is shown as a solid line in 
Fig.\ref{fig:AllEurope_cpd}, and is given by
\be
\frac{d{\P}}{dt} \Bigg{/} \frac{{\P}_{max}}{e\tau} = 
\frac{ e^{-((t-t_0)/\tau\, - 1)} }{\exp (e^{-(t-t_0)/\tau})} \ .
\label{eq:dPdt_univ_infig}
\ee

\begin{figure}[t]
\includegraphics[width=0.9\linewidth]{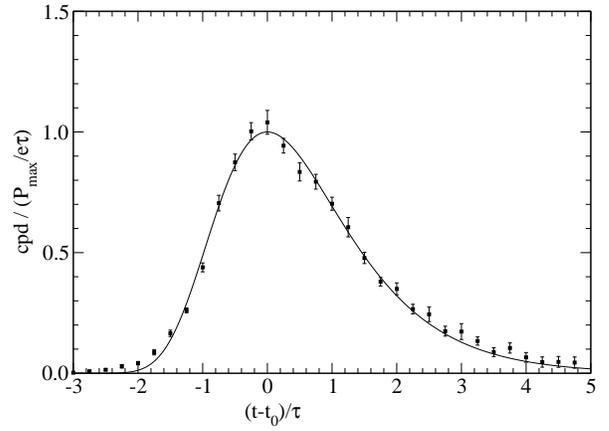}
\caption{A country-averaged curve for $d{\P}/dt$ obtained by binning the 15 datasets in 
Fig.\ref{fig:AllEurope_cpd} and averaging within each time bin.
The resulting $d{\P}/dt$ values are shown together with the 
parameter-free SHIR model prediction (\ref{eq:dPdt_univ_infig}).}
\label{fig:dPdt_binned_allavg}
\end{figure}

We can use this scatterplot to combine the data from the 15 European countries we have 
considered, since they have now been superimposed through translating in time and scaling 
both axes. On binning in the scaled time variable $(t-t_0)/\tau$, we can calculate the 
average and variance of the scaled cpd datapoints in each bin, which provides an average 
and error for $d{\P}/dt$ at that value of $(t-t_0)/\tau$. As an example, we have used bins 
of width $\Delta (t-t_0)/\tau = 0.5$, centered on $t-t_0 = 0$, which gives the result 
shown in Fig.\ref{fig:dPdt_binned_allavg}. This appears approximately consistent with 
the parameter-free SHIR model result for $d{\P}/dt$ (\ref{eq:dPdt_univ_infig}), although 
there may be relatively minor discrepancies at onset ($(t-t_0)/\tau \approx -2$) 
and well after peak ($(t-t_0)/\tau \approx 3-4$).

The Europe-averaged COVID-19 data for the new cases per day shown as points with errors in 
Fig.\ref{fig:dPdt_binned_allavg} can be used to calculate expectation values under this
common COVID-19 $d{\P}/dt$ distribution. Some of these results are 
given in Table.\ref{table:dPdt_expected_values}, where they are compared to
theoretical values predicted by the SHIR model $d{\P}/dt$, Eq.(\ref{eq:dPdt_univ_infig}). 
Evidently the simpler expected values are in reasonably good agreement, 
notably the displaced value of the mean from the maximum. There is however a discrepancy 
in the cubic expected values, which are sensitive to the relatively less well determined 
wings of the distribution at larger $|t-t_0 |/\tau$.

\begin{table}[h]
\begin{center}
\begin{tabular}{|c|c|c||l|}
\hline
\ & & & 
\\
Quantity
& SHIR theory 
& SHIR num.
& COVID data
\\
\hline
\ & & & 
\\
\
Area
& e
& 2.718\phantom{0} 
& \phantom{00}2.787 (25)
\\
\
$x$
& $\gamma$
& 0.5772
& \phantom{00}0.5746 (28)
\\
\
$x^2$
& $\gamma^2 + \pi^2/6$ 
& 1.978\phantom{0}
& \phantom{00}1.996 (52)
\\
\
$\sigma$ 
& $\pi/\sqrt{6}$
& 1.283\phantom{0}
& \phantom{00}1.291 (8)
\\
\
$x^3$
& $\gamma^3 + \gamma \pi^2/2 + 2\,\zeta(3)$
& 5.44\phantom{00}
& \phantom{00}4.57 (9)
\\
\
skewness
& $12\sqrt{6}\,\zeta(3) / \pi^3$
& 1.14\phantom{00}
& \phantom{00}0.70 (19) 
\\
\hline
\end{tabular}
\vskip 2mm
\caption{Expected values of several quantities calculated using the theoretical
SHIR model cases per day distribution function $d{\P}/dt$ (\ref{eq:dPdt_univ_infig}), compared 
to the results found using the Europe-averaged COVID-19 data shown in Fig.\ref{fig:dPdt_binned_allavg}.
Here for simplicity we abbreviate $(t-t_0)/\tau = x$.}
\label{table:dPdt_expected_values}
\end{center}
\end{table}

\eject

\subsection{Future prospects}

We have shown that a SHIR (generalized SIR) model and its associated differential equations give rather good
closed-form results for the COVID-19 data on new cases per day and cumulative cases, when solved under 
certain assumptions. These assumptions can be relaxed in future studies. 

One important assumption was to ignore the effects of the
``resolved" population fraction ${f_{\R}}$. This quantity, which is certainly of interest in planning
economic recovery, is simple to evaluate from the ``infected" fraction we have discussed here; 
it is simply the integral in Eq.\ref{eq:SHIR_DE3_soln}. This can be evaluated analytically,
together with the {\it full} DE for ${f_{\I}}$, including the $r_{\R}$ term, to accommodate 
transitions out of ``infected" and into ``resolved." The cumulative case numbers ${\P}(t)$ 
discussed in relation to COVID-19 data should then be defined by 
${\P}(t) = {\N}_0 (f_{\I} + f_{\R})$, 
{\it i.e.} ``infected" plus ``resolved," since $f_{\R}$ is no longer assumed to be negligible.

\begin{figure}[ht]
\includegraphics[width=0.9\linewidth]{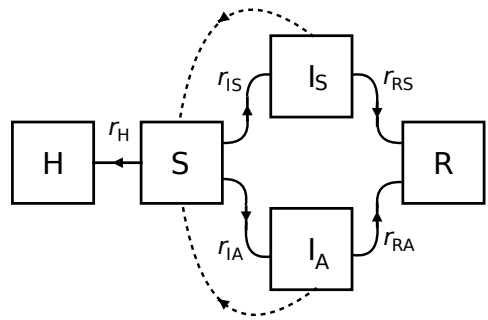}
\vskip -2mm
\caption{An extended SHIR model block diagram including an asymptomatic line.}
\label{fig:SHIR_blockdiag2}
\end{figure}

Calculations of mortality rates are certainly of interest, and are implicitly included in the transition from
the infected population ${\I}$ (or ${\I}_{\S}$) to the resolved population ${\R}$. Mortality can easily 
be treated separately, for example by specifying separate ${\I} \to {\R}$ rate parameters for mortality and 
for recovery. In the extended SHIR model of Fig.\ref{fig:SHIR_blockdiag2}, mortality would presumably 
only contribute to the combined rate for ${\I}_{\S} \to {\R}$. The mortality rate coefficients may of course 
depend strongly on the procedures followed at each country's medical facilities.

Numerical methods can be used to extract SHIR model results for the full model, including the 
${\S}-{\I}$ quadratic infection term, and can test the importance of the ``small fraction of infections" 
approximation used here. 

A straightforward generalization of the model could accommodate the effect of ``asymptomatic" 
infected individuals on the dynamics of the COVID-19 pandemic. 
Recent antibody results from Italy \cite{Lav20} suggest that the 
asymptomatic cases are a large but not dominant fraction of all COVID-19 cases.   
Once better data on the number of asymptomatic cases becomes available, 
it may prove useful to extend the model to include this population. This group follows 
a progression ${\S} \to {\I}_{\A} \to {\R}$, and while in the ${\I}_{\A}$ population 
presumably contributes to the infection of other individuals still in ${\S}$.
Although adding this branch to the model may indicate that the fraction of resolved individuals
${f_{\R}}$ is much larger than the SHIR model alone suggests, the pathway through symptomatic cases
will remain as before. The most visible predictions of the model, for the number of symptomatic cases 
(now ${\I}_{\S}$), may not change significantly. A block diagram of this extended SHIR model is shown in 
Fig.\ref{fig:SHIR_blockdiag2}. 

Finally, although we have primarily considered COVID-19 in Central and Western European countries,
largely because of an assumed similarity in standards for reporting data, there is extensive data 
available from many other countries which could be addressed using this type of model, 
undoubtedly with interesting results.

\section{Summary and Conclusions}

In this paper we have introduced a generalization of the SIR model 
in order to incorporate the effects of pandemic reduction measures. 
This ``SHIR" model adds a rate process that allows transitions from the 
susceptible population (S) to a new ``hidden" population (H), which we assume is
not susceptible to infection. We have solved this model analytically in the limit of a small
infected population fraction, and showed that the predicted time dependence of 
cumulative cases and new cases per day is in surprisingly good agreement with COVID-19 data. 
(We primarily considered Central and Western European countries in this study.) 
The solution of the model with our approximations involves three free parameters; 
one is the ``social response time" $\tau$, which varies widely between countries, 
in a manner that correlates with their coronavirus policies.

\section{Acknowledgments}

We are happy to acknowledge useful communications with and suggestions from
P.Ardanuy, D.Arovas, F.E.Close, E.Kioski, S.Nagler, M.Osmann, J.-L.Rosenthal, M.Savage, M.Shinn, D.J.Smith, 
D.A.Spong and E.S.Swanson during the production of this manuscript. We are also grateful to E.S.Swanson 
for assistance with calculations of expected values and errors in the SHIR model, and with the model block diagrams.

\vfill\eject


\begin{thebibliography}{99}

\bibitem{Roc2014} 
K. Rock, S. Brand, J. Moir and M.J.Keeling,
``Dynamics of infectious diseases"
Rep.\ Prog.\ Phys. {77}, 026602 (2014).
http://doi.org/10.1088/0034-4885/77/2/026602

\bibitem{fI_max} In this simple SIR model the time at which $f_{\I}$ is a maximum and 
that maximum value can be solved for analytically, and are respectively 
$t = \ln(r_{\I}/r_{\R})/(r_{\I}-r_{\R})$ and
$f_{\I} = (\rho / (\rho - 1))( \rho^{-1/(\rho-1)} - \rho^{-\rho/(\rho-1)})$,
where $\rho = r_{\I}/r_{\R}$.

\bibitem{Ker1927} 
W. O. Kermack and A. G. McKendrick, 
``A contribution to the mathematical theory of epidemics"
Proc.\ R.\ Soc.\ Lond.\ {A115}, 700 (1927).
http://doi.org/10.1098/rspa.1927.0118.

\bibitem{Hbox}
Protected groups similar to the ``H" box of Fig.\ref{fig:SHIR_blockdiag1}
have previously been considered in generalizations of the SIR model. 
For example, the effects of vaccination are discussed using an SIRV model \cite{Roc2014}, and the
2003 SARS model of Lipsitch {\it et al.} \cite{Lipsitch2003} introduces an XQ population to model quarantine.
 
\bibitem{Lipsitch2003}
M. Lipsitch {\it et al.}, 
``Transmission Dynamics and Control of Severe Acute Respiratory Syndrome."
Science 3000 (5627), 1966-1970; doi: 10:1126/science.1086616.

\bibitem{Ard}
The use of ``resolved" for the post-infected population ${\R}$, the sum of deceased and recovered, 
was suggested by P.Ardaury. This population is also referred to as ``closed cases" in the Worldometer 
COVID-19 database \cite{WCU}.

\bibitem{WCU} ``Worldometer Coronavirus Updates''
URL: www. worldometers. info / coronavirus / countries .

\bibitem{Gom1832}
B. Gompertz, ``On the Nature of the Function Expressive of the Law of Human Mortality, and on 
a New Mode of Determining the Value of Life Contingencies.'' Phil. Trans. Roy. Soc. London 123, 
513-585, (1832). 

\bibitem{Wei}
Eric W. Weisstein, ``Gompertz Curve." 
From MathWorld -- A Wolfram Web Resource. https:// mathworld. wolfram. com / GompertzCurve.html .

\bibitem{def_skewness} Skewness is a dimensionless property of a distribution, 
and is defined by $<(t-<t>)^3> / \sigma^3$, where the averages here are taken using 
the distribution $d{\P}/dt$. 

\bibitem{AusNorWCUpts}
We note in passing that a single ``outlier" point in the Austrian cpd data, 
1321 new cases on 26 Mar 2020, lies above the upper cutoff of 1000 cases per day in 
Fig.\ref{fig:Austria_cpd_SHIRfit}. It was of course
included in the fit. Similarly, the Norwegian cpd data includes 
an outlier of 399 new cases on 27 Mar 2020 and
a negative entry of -44 new cases on 08 Apr 2020, 
which we assume was a correction to the data. 
These points were included in the Norwegian fit shown in 
Fig.(\ref{fig:Norway_cpd_twopeaks}).

\bibitem{cas_fit}
It is notable that the estimated parameter errors in the cumulative ``cas" data fit are much smaller than in
the preceding new cases per day ``cpd" fit. The cas parameter errors are likely underestimated, since the 
cas data points represent a running sum, and are therefore highly correlated. The cpd parameter errors are 
presumably more realistic, since they are not correlated in this manner. Note that the differences between the 
fitted cpd and cas parameters are comparable to the cpd errors.

\bibitem{Sweden_policy}
M. Paterlini, Nature {580}, 574 (2020).
doi: 10.1038/ d41586-020-01098-x .

\bibitem{Netherlands_policy}
Prime Minister M. Rutte, National Address (16 Mar 2020),
https:// www.rijksoverheid.nl/ documenten/ toespraken/ 2020/03/16/
tv-toespraak-van-minister-president-mark-rutte.
English translation, https:// nltimes.nl/ 2020/03/16/ 
coronavirus-full-text-prime-minister-ruttes-national-address.

\bibitem{UK_policy}
British Prime Minister Boris Johnson Coronavirus Address (23 Mar 2020).
https:// www.c-span.org/ video/ ?470622-1/
british-prime-minister-announces-lockdown-televised-address.

\bibitem{Lav20}
Lavezzo, E. {\it et al.}, 
``Suppression of a SARS-CoV-2 outbreak in the Italian municipality of Vo'.'' 
Nature https:// doi.org/10.1038/s41586-020-2488-1 (2020).

\end{thebibliography}
\end{document}